\documentclass{article}
\usepackage[utf8]{inputenc}
\usepackage[colorlinks=true,linkcolor=blue,citecolor = blue]{hyperref}%

\usepackage{amsmath}
\usepackage{amsthm}
\usepackage{arxiv}
\usepackage{natbib} %[sort, numbers]
\usepackage{graphicx}
\usepackage{xr}
\usepackage[dvipsnames]{xcolor}
\colorlet{Mycolor1}{green!10!orange!90!}
\colorlet{Mycolor2}{green!30!black!70!}
\colorlet{Mycolor3}{blue!30!red!100!}
\usepackage{textcomp}
\usepackage{multirow}
\usepackage{booktabs}
\usepackage{adjustbox}
\usepackage{amsmath}
\usepackage{threeparttable}
\usepackage{amssymb}
\usepackage{enumitem}

\usepackage{algorithm}
\usepackage[noend]{algpseudocode}

\makeatletter

\theoremstyle{definition}

\theoremstyle{remark}
\newtheorem{remark}{Remark}

\title{A Sparsity Algorithm with Applications to Corporate Credit Rating}
\author{Dan Wang, Zhi Chen, Ionut Florescu }
\date{October 2020}

\begin{document}

\maketitle

% \begin{abstract}
 
%     This work defines the problem of finding a counterfactual explanation as an optimization problem. A sparsity algorithm is proposed which solves the problem while also maximizing the sparsity of the solution. We apply the algorithm to public companies in order to improve their credit ratings. We validate the algorithm with a synthetically generated dataset and further apply to quarterly financial statements. The evidence indicates counterfactual explanation can capture the real statement features that changed between the current quarter and the following quarter when ratings improved. The empirical results show that the ``effort'' required to improve credit rating is positively associated to the credit rating level.
% \end{abstract}

\begin{abstract}
    In Artificial Intelligence, interpreting the results of a Machine Learning technique often termed as a black box is a difficult task. A counterfactual explanation of a particular ``black box'' attempts to find the smallest change to the input values that modifies the prediction to a particular output, other than the original one. In this work we formulate the problem of finding a counterfactual explanation as an optimization problem. We propose a new ``sparsity algorithm'' which solves the optimization problem, while also maximizing the sparsity of the counterfactual explanation. We apply the sparsity algorithm to provide a simple suggestion to publicly traded companies in order to improve their credit ratings. We validate the sparsity algorithm with a synthetically generated dataset and we further apply it to quarterly financial statements from companies in financial, healthcare and IT sectors of the US market. We provide evidence that the counterfactual explanation can capture the nature of the real statement features that changed between the current quarter and the following quarter when ratings improved. The empirical results show that the higher the rating of a company the greater the ``effort'' required to further improve credit rating.
\end{abstract}

\section{Introduction}

Corporate credit rating is an assessment of the credit risk level of a company. Generally, it is issued by a credit rating agency such as Standard and Poor, Moody's, or Fitch \citep{ratings2018guide}. 
\citep{spguidecr}. 
The rating expresses the agency's opinion about a company's ability to meet its financial obligation in full and on time. This rating serves as an aid to financial investors in order to assess various investment opportunities. Since the credit rating is supposed to be a uniform measure across companies, it enables investors to compare risk levels of companies which issue the financial instruments present in their portfolios.

Thus, credit rating is a very important measure for companies. Credit rating expedites the process of purchasing and issuing bonds by providing an uniform and efficient measure of credit risk  \citep{akdemir2012assessment}. Thus, instead of borrowing loans from banks, public companies are more likely to raise money from capital markets by issuing bonds and notes. A good credit rating is beneficial to companies. Bond yields are negatively related to credit rating \citep{luo2019bond}. That is, a higher credit rating can help public companies raise funds for a lower repayment cost. Further, a good credit rating means that the company is less likely to default on their obligations, thus attracting risk-averse investors such as pension funds and mutual funds \citep{dittrich2007credit}.

Recent literature is implementing machine learning and deep learning techniques to assess public corporations' credit rating. In the US markets, corporate credit rating has been evaluated using Support Vector Machines (SVM's), Tree based models and network learning methods such as Artificial Neural Network (ANN), Convolutional Neural Network (CNN) and Long short Term Memory (LSTM) \citep{ye2008multiclass, wallis2019credit, hajek2011credit, hajek2014predicting,golbayani2020application,wang2020image}. 
Deep learning techniques are popular to assess credit risk in European market and Asian market as well \citep{khashman2010neural, kim2010support, west2000neural, khemakhem2015credit, zhao2015investigation,addo2018credit}.

Even though machine learning and deep learning methods have achieved considerable accuracy for various types of classification problems \citep{lecun2015deep}, and in particular for credit risk assessment \citep{golbayani2020comparative}, the constructed neural networks continue to be treated as a black-box method. This black-box maps the input features into a classification output without low-level explanation \citep{chakraborty2017interpretability, carvalho2019machine}. 
However, it is very important for a model to provide visibility and interpretability. Which specific features are important and how these features impact the output. In finance, interpretability of the model is critical, as it is required by law. Financial regulation provides investors the right to receive an explanation of the algorithms used by investment firms \citep{regulation2018general,regulation2016regulation,goodman2017european}.

Interpretable machine learning is a fast growing field that addresses this issue. It is defined as the use of machine learning or deep learning models to extract relevant knowledge about domain relations contained in the data \citep{murdoch2019interpretable}. The interpretability of a machine learning method may be divided into: (1) model explanation, and (2) post-hoc explanation. Model explanation means that the model is inherently interpretable and can generate explanations when trained \citep{yang2016hierarchical}. The post-hoc explanation, refers to the capability of the model to generate an explanation based on existing decisions \citep{mordvintsev2015inceptionism, plumb2018model}. 

% \textbf{In finance, interpretable machine learning is widely adopted when build a model with interpretability. - this makes 0 sense} For example, \cite{bussmann2021explainable} propose a boost machine learning model that not only can maintain the high accuracy but also interpret the predictive output.

%\dan{Even though the traditional machine learning models are capable of explaining why and how a decision is made by a existing model and data, they are not able to answer question like what the data would be look like when the decision is changed. } \textbf{here I want to say that ML models can explain why this happen when x is given, but they cannot answer the question about when y is change, what x should be look like. This will introduce our counterfactual explanation model}

In mathematics once a functional relationship $f$ between an input $x$ and an output $y$ is constructed\footnote{i.e., a deterministic function},  one is able to determine the preimage of a set. That is, calculate $f^{-1}(B)=\{x : f(x)\in B\} $ defined as the set of inputs $x$ that take values in $B$, where $B$ is a set in the co-domain of $f$. Generally speaking, a machine learning technique creates a relationship $x\xrightarrow{f} y$, where $f$ is provided by the ML technique used, and typically $y$ is categorical. Thus, given another input $x'$, a constructed ML technique is able to calculate the output of $x'$ by simply calculating $f(x')$.  However, providing the preimage of a specific category is hard. Often, the domain set is ill defined, and furthermore there is randomness in the ML technique. For instance, if a particular $x$ has a 50/50 probability of being in category $y_1$ respectively $y_2$, then should the preimage of $y_1$ contain $x$ or rather should $x$ be in the preimage of $y_2$? 

In order to answer such questions, the counterfactual explanation was introduced for ML techniques \citep{wachter2017counterfactual}. Specifically, given a ML technique $f$ that associates an input $x$ with an output $y$, if we want the value of $f$ to be $y'$ how should the particular input $x$ be modified so that the value of the modified $x$ is $y'$? The counterfactual explanation technique was applied to image recognition, healthcare and language models  \citep{goyal2019counterfactual, huang2019reducing, prosperi2020causal}. The closest application of counterfactual explanation to finance we could find is to credit cards application with a binary black box classifier \cite{grath2018interpretable}. Specifically, in the paper cited, the author use the counterfactual explanation to provide advice about how to change applications for credit cards in order to have a successful outcome. The authors solve a typical optimization problem using a Median Absolute Deviation (MAD) norm. In our work we are modifying the optimization problem by focusing on the sparsity of the counterfactual solution. 

% % \textbf{this next makes 0 sense again. If i have a NN created I just put the new input in and I get y. Please read what I wrote}
% % ``What would happen to the decision if  we had different features as input?''.show that counterfactual explanation can well explain the prediction of credit application with black-box classifiers.
% This work is motivated by counterfactual explanation \citep{wachter2017counterfactual}. A counterfactual explanation describes a generic causal situation in the form:
% \textit{Score $y$ was returned because variables $X$ had values $(x_1, x_2, \dots)$ associated with them. If $X$ instead had values $(x_1', x_2', \dots)$, and all other variables had remained constant, score $y'$ would have been returned.}
% wtf this is obvious. where did you get this? \dan{it is defined in \citep{wachter2017counterfactual} and \cite{grath2018interpretable}, I believe it make more sense if we mention that $y'$ is desired output}

\begin{figure}[htbp]
\centerline{\includegraphics[width=10cm]{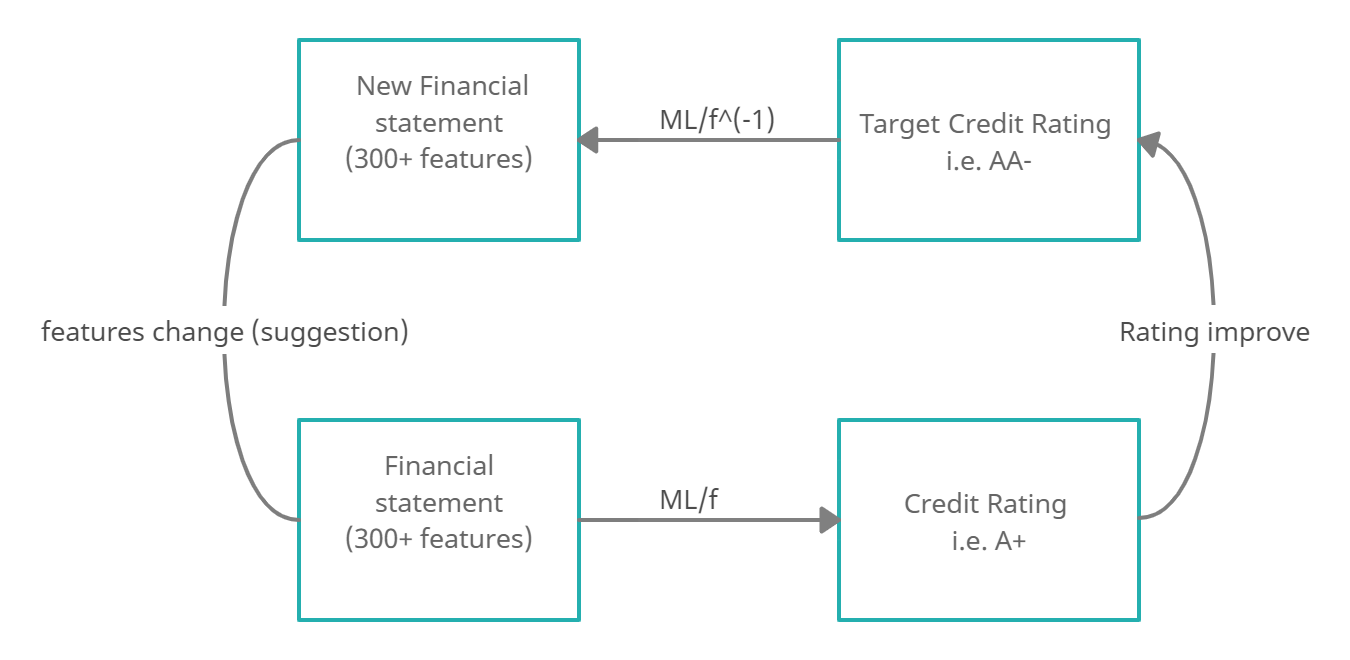}}
\caption{Example of counterfactual explanation\label{example}}
\end{figure}

We next discuss the challenges faced when calculating a counterfactual explanation. In a classification problem a high dimensional input $x\in \mathbb R^n$ is assigned through $f$ to a $y$ in a countable (often finite) set of outputs. Therefore, the ML ``function'' is not injective. This means that for each output $y$ there is a range of inputs $x$ which are mapped into it.  Now look at Fig.\ref{example} which describes our credit rating problem. Obviously, the solution $f^{-1}(\text{Target rating})=f^{-1}(y')$ is not unique. That is, there are multiple $x'$'s which will map into the new $y'$. To make matters even more complex the function $f$ is in fact ``probabilistic''. For example, the same input $x$ is associated to output $y_1$ with probability $p_1$, and with output $y_2$ with probability $p_2$. Since $p_1>p_2$ the ML decided $f(x)=y_1$ but in fact, the magnitude of the probabilities needs to be taken into consideration as well. 

%As shown in , the objective of this work is to provide a suggestion for a company to help it improving credit rating.
When studying corporate credit rating, there are two considerations worth mentioning. First, a financial statement contains a large number of features. For instance, the Compustat\textregistered{} dataset \citep{SP2019compustat}, contains $332$ financial accounting variables collected from the original quarterly financial statements. It is impractical for a company to focus on changing all $332$ features. Second, some of the features may not be possible to change, for example: Comprehensive Income - Noncontrolling Interest, Equity in Earnings (I/S) - Unconsolidated Subsidiaries. 

The purpose of this paper is to set up a proper optimization problem to address these issues. The central idea is to  minimize the number of features modified for $\delta_x$, where $x+\delta_x$ is the counterfactual explanation. In finance, particularly when applied to corporate credit rating, this allows company's decision makers to focus their attention when trying to improve company's credit rating. In section \ref{method} we describe the optimization problem and propose an algorithm we call ``the sparsity algorithm'' to solve it. Section \ref{result} presents experiment results obtained from both simulated data and real rating data.

%Good credit ratings allow companies to easily borrow from financial institutions or public debt markets. For credit rating interpretable problem, the objective is to provide a recommendation for public companies to get a better credit rating.

%It is easy to compute counterfactual input values $(x_1', x_2', \dots)$ for a desired $y'$, where $y'$ is a better credit rating comparing with current rating. However, Giving recommendations to a company with all inputs changed seems impractical. In this paper, we propose a model to improve the credit rating for a company by only change small number of features.

\section{Methodology}\label{method}

In this section, we describe the optimization problem and the algorithm introduced to solve this problem.

\subsection{Statement of problem}\label{ori_problem}
As described in the introduction, the goal of this work is to discover the smallest subset of input data that can realistically be changed, so that the output of the model is reclassified for this changed input. To achieve this goal we propose solving a minimization problem. 

Specifically, given a trained deep learning model $f(\cdot)$  which relates input variable $x$ with a specific classification $y$, the problem is to find $\delta_x$ such that the response of $x+\delta_x$ is a different class $y'=f(x+\delta_x)$ than the response for $f(x)=y$. However, only certain components of $x$ can realistically be modified. Further, the problem attempts to find the smallest modification of $x$ which will accomplish the respective reclassification. Thus, we minimize the L0 ``norm'' of $\delta_x$, and we impose a mask $w$. The problem to solve is expressed mathematically as: 

\begin{equation}\label{eq: origin}
\begin{aligned}
    & \min_{\delta_x} \quad  \|\delta_x\|_0 \\
    \textrm{s.t.} & \ f(x + w \circ \delta_x) \leq y'.\\
\end{aligned}
\end{equation}

Here, $\|\cdot\|_0$ is a mapping which counts the nonzero elements of a vector. This operator, often described as the L0 norm, is not actually a norm. However, it is used extensively in the Machine Learning area \citep{shukla2018machine}. $\circ$ is the Hadamard product that calculates the product element-wise of two matrices of the same dimensions. $w$ is a predefined vector with values 0 and 1, which masks the input components of $x$ which are not modifiable.  $y'$ is the desired output.

With $\widehat{\delta_x}$ denoting the solution of the problem \eqref{eq: origin}, the counterfactual explanation is  $x'= x + \widehat{\delta_x}$ \citep{wachter2017counterfactual}. In a credit rating problem, $y'$ is typically taken as a one grade upgrade from the original credit rating $y$, but in principle it may be any target rating. 

\begin{remark}
In this paper $f$ is modeled using a Multi-Layer Perceptron (MLP). MLP is the most prevalent network architecture for credit rating problems \cite{ahn2011corporate,huang2004credit,kumar2003forecasting, kumar2006artificial}. In credit rating applications the output layer of $f$ contains the distinct classes of the corporate credit rating. The error in prediction is obtained by applying a categorical cross-entropy loss function on the output layer.  A \textit{GridSearch} has been applied to find the optimal values of the MLP hyper parameters for our specific datasets. 
\end{remark}

%$f(\cdot)$ is machine/deep learning model, $\lambda$ is the weights balances the counterfactual between obtaining the exact desired output and making the smallest possible changes to the input vector $x$. 

\subsection{The Algorithm}

The sparsity minimization problem is a well studied problem \citep{yuan2016sparsity, cai2013exact, zhang2020top}. However, the problem as written in \eqref{eq: origin} is still very challenging when the function $f$ is complicated as is the case of a deep neural network. The difficulty comes from L0 not being a norm, as well as from the function $f$ in the problem having a very complex form. In previous work \cite{grath2018interpretable}, the authors use Median Absolute Deviation (MAD) in order to impose sparsity. In our case their approach is not feasible for two reasons. First, MAD imposes sparsity by minimizing the size of the change of certain features (the features that are far from the median). In practice, this results in changing ALL components with some components having relatively small changes. For our finance applications, sparsity means that most components of the change have to be exactly $0$. Second, the MAD weights used in the optimization are determined automatically from the dataset. In our application, some features cannot be modified. Thus, we need to define the problem in a way that will allow the algorithm to only modify pre-specified features. 

In our approach to solve \eqref{eq: origin}, we replace $L0$ with the $L_1$ norm. There are two reasons for this. First, the $L_1$ norm has been previously used as a regularizer, to increase sparsity \citep{bruckstein2009sparse, selesnick2017sparse}. Second, since it is a proper norm, we can rewrite the problem \eqref{eq: origin} as the following unconstrained optimization problem. 

\begin{equation}\label{eq: mod0}
\begin{aligned}
    \min_{\delta_x} \quad &  \lambda(f(x + w \circ \delta_x) - y')^2  + ||\delta_x||_1\\
\end{aligned}
\end{equation}

Note that problem \eqref{eq: mod0} treats the output of $x+w \circ \delta_x$ as a single number. However, as mentioned in the introduction, most Machine Learning methods take the decision based on a likelihood set of probabilities associated to each of the discrete $y$ outputs. To handle this issue, $f(\cdot)$ is replaced with the set of probabilities denoted $F(\cdot)$ (the output distribution). The output $y'$ is replaced with the ideal probability set  $Y'$ \citep{janocha2017loss}. We thus replace the first part of the loss function in equation \eqref{eq: mod0} with the cross-entropy \citep{kline2005revisiting} in equation \eqref{eq: mod1}. In this way, we can inform the managers how their entire  credit rating  probability distribution will be modified following the algorithm's recommendation.

\begin{equation}\label{eq: mod1}
\begin{aligned}
    \min_{\delta_x} \quad &  -\frac{\lambda}{N} \sum^N (Y' \cdot log(F(x + w \circ \delta_x)) )  + ||\delta_x||_1\\
\end{aligned}
\end{equation}

Since the problem is now unconstrained we can use the gradient descent method to solve this problem. Gradient descent is a good way to solve such optimization problems when the objective function is convex and differentiable \citep{cauchy1847methode, curry1944method}.

However, the solution of the unconstrained problem \eqref{eq: mod1} is not necessarily sparse and also cannot guarantee that $f(x + w \circ \delta_x) \leq y'$. This means that the counterfactual solution may not always produce a better credit rating. To solve these issues, we propose a new algorithm as follows. 

\begin{algorithm}
\caption{Sparsity algorithm}\label{alg:euclid}
\begin{algorithmic}[1]
\State  Define the masking variable $w$ by an accounting expert or the client company. 
\State  Solve equation \eqref{eq: mod1} using the gradient descent to get $\delta_x$. Due to the masking variable $w$, the  $\delta_x$ only has $n$ nonzero coordinates which correspond exactly to the values of $1$ in the $w$. 
\State $n$ is generally too large and we want to focus on a small number of changes. Let $k$ denote the practical number of changes that may be implemented. Let $$\delta_x'=\left|\frac{\delta_x}{x}\right|,$$ denote the absolute magnitude of change relative to the original vector $x$.
\State {We set $\delta_{x_i}' = 1$ when the component $x_i = 0$.}
\State We construct vectors $\delta_{x,1}, \delta_{x,2}, \dots, \delta_{x,k}$ in the following way. For each $i$ in $1,2,\dots,k$, $\delta_{x,i}$ holds the values in $\delta_x$ which correspond to the $i$ largest values in $\delta_x'$. All other components in $\delta_{x,i}$ are set to 0.
\State Set t = 1
\While {$ f(x + w \circ \delta_{x,1}) > y'$ \& $ f(x + w \circ \delta_{x,2}) > y'$ \& \dots \& $ f(x + w \circ \delta_{x,k}) > y'$ \& $t < T$}
\State increase $\lambda$ by a predefined proportion
\State Do steps 2 to 4
\State t = t + 1
\EndWhile\label{euclidendwhile}
\State \textbf{return} $\delta_{x,i}$, where $f(x + w \circ \delta_{x,i}) \leq y' $ 

\end{algorithmic}
\end{algorithm}

Comparing with the solution $\delta_x$  obtained directly from equation \eqref{eq: mod1}, the sparsity algorithm produces a vector with a large number of $0$ components (a sparse vector). The algorithm accomplishes this task by following three main steps. First, the algorithm calculates the change ratio $\delta_x'$ for each element in the output vector $\delta_x$ relative to the original vector. Second, it constructs $k$ candidate vectors, going less sparse from vector $1$ to vector $k$. The algorithm repeatedly solves the problem by putting more and more importance on the boundary condition ($f(x + w \circ \delta_x)\leq y'$). We end the process if there is at least one candidate solution which qualifies the input for the better rating $y'$. If there is no solution to the sparsity algorithm, we interpret it as the rating may not be changed in a simple way for the given input vector $x$.

In the last loop there are two issues when returning the final value $\delta_{x,i}$. 
\begin{remark}
If the $i$ in not unique and there are multiple solutions for example $\delta_{x,i}$ and $\delta_{x,j}$, we can choose the final output based on whether $i$ or $j$ achieve a smaller value in equation \eqref{eq: mod1} or we can chose the variant with the smallest number of nonzero coordinates. 
\end{remark}

\begin{remark}
If we reach step $T$ and there is no solution then the rating of the company cannot be improved based on the existing credit rating model. 
\end{remark}

\begin{remark}[Step 4 in the algorithm]\label{step4}
When a component of $x$ equals $0$ the relative change $\delta_x'$  in step 3 for that component will be infinite. This forces the sparsity algorithm to favor choosing the features with values equal to $0$. This step in the sparsity algorithm is introduced to set a ceiling for the ratio in order to resolve this issue. However, we will  discuss another possible solution in Section \ref{exp2:real_rating} when we apply the algorithm to financial data. 
\end{remark}

\subsection{What is the practical importance of the sparsity algorithm? }\label{why_sig}

The idea of this work is simple. Given a learned algorithm $f(\cdot)$, which associates a categorical (rating) $y$ to an input $x$, can we find $\delta_x$ a change in $x$ so that the new rating associated to the changed $x+w\circ \delta_x$ (counterfactual) is now $y'$? In this context, we call the distance ($\delta_x$) between the original input $x$ and the counterfactual input as \textbf{effort}. In the context of credit rating this $\delta_x$ calculates how much actual effort has to be put in changing the qualified rating. Having a sparser solution $\delta_x$, may translate into a smaller effort to change $x$ while making sure that the output class of $x + w\circ \delta_x$ has been improved by at least one notch.

\section{Empirical Results} \label{result}

We will be using two sets of data to demonstrate the validity of the proposed algorithm. First, we shall use synthetically generated data to illustrate the performance of the algorithm on a simple to understand case. Second, we use quarterly fundamental data obtained from the Compustat Database \citep{SP2019compustat}. The fundamental data contains 332 accounting variables including balance sheet data, income statement data, etc. We use Standard and Poor's credit ratings as the target rating $y$. The case study 2 is the real financial study we wish to analyze.

In this work we are interested in answering three different questions. 
\begin{enumerate}
    \item \textit{Are the results of the sparsity algorithm intuitively correct when using the synthetically generated data?}
    \item \textit{Is it possible to improve credit rating with less effort then it actually happened in reality?}
    \item \textit{Does the effort to improve rating depend on the rating? Specifically, do we need to exert more effort when changing rating from non-investment grade to investment grade than to change rating within the investment grade?}
\end{enumerate}

\subsection{Case study: synthetically generated data}

Since $f(\cdot)$ is the solution to a machine learning algorithm, in principle we could attempt to prove mathematically that the sparsity algorithm can solve the equation \eqref{eq: origin}. The Lagrange multiplier version of the problem in equation \eqref{eq: mod1} is well posed and the gradient descent will provide the optimal solution. The sparsity algorithm imposes constraints on the solution and it fundamentally is checking how close the solution is to the original problem \eqref{eq: origin}. Thus the mathematical proof idea is to show that the sparsity algorithm produces an improvement at every step and that in the limit we obtain the solution of \eqref{eq: origin}. 

However, such proof would be dry and would only bring joy to mathematically inclined. We chose to follow a different approach. In this section we design an intuitive case study by synthetically generating data in such a way that would have an easy to understand solution. We compare the solutions obtained using the classical gradient descent and the solutions obtained using the sparsity algorithm. We perform matched pairs one-sided t tests on the L0 and $L_1$ norms to compare these solutions.

%\subsubsection{describe the data} \label{sim_data}

We create a $5$-dimensional dataset with points $X = (X_1, X_2, X_3, X_4, X_5)$, where all of the features are normally distributed random variables. We let $X_1$ and $X_2$ denote the important variables and we let $X_3$, $X_4$, and $X_5$ be noise variables. Specifically, the $X_1$ and $X_2$ variables are each a mixture of normals with means $1$ and $-1$ and variance $0.3$. Their pdf is: 
$$ f_{X_1}(x)= f_{X_2}(x)=\frac 12 \frac{1}{\sqrt{2\pi (0.3)}}e^{-\frac{(x-1)^2}{2(0.3)}}+ \frac 12 \frac{1}{\sqrt{2\pi (0.3)}}e^{-\frac{(x+1)^2}{2 (0.3)}}$$
The $X_3$, $X_4$ and $X_5$ are iid normally distributed with mean $0$ and variance $0.3$.  

\begin{figure}
    \centering
    \includegraphics[width=0.7\textwidth]{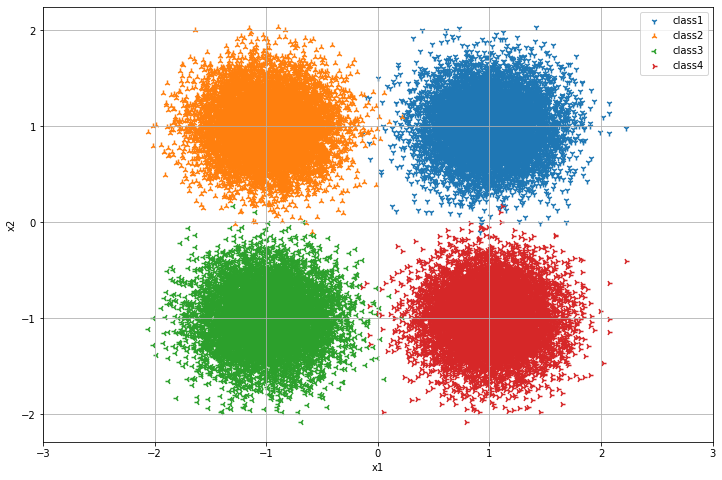}
    \caption{Data visualization for $x_1$ and $x_2$}
    \label{fig:x1x2}
\end{figure}

Figure \ref{fig:x1x2} shows the projection of the synthetically generated points $X$ on the first two coordinates. We can clearly see the centers of the $4$ classes. In this synthetically generated data, we arbitrarily define blue points as rating 1, orange points as rating 2, green points as rating 3, and red points as rating 4 (counterclockwise starting from the first quadrant). We make the convention that ratings $1$ is the best, decreasing with $4$ being the worst. In this experiment, we aim to improve the rating of the points using the smallest effort $\delta_x$.

The point of this synthetically generated case study is to showcase the results of the algorithms in a context where we can plot and actually see the results.

\subsubsection{Results obtained when using the synthetically generated data}

\begin{figure}
    \centering
    \includegraphics[width=0.7\textwidth]{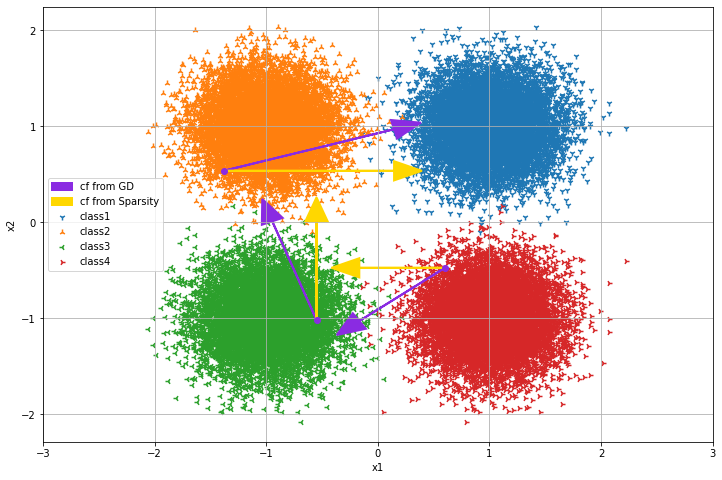}
    \caption{Algorithm applied on synthetically generated data}
    \label{fig:exp1}
\end{figure}

As mentioned, we want to determine which coordinates need to be changed to ``improve the rating''. In this simple exercise, for a point $x$ this translates into determining the ``best'' $\delta_x$ that will improve the class number. To illustrate the performance of the algorithm we pick 3 points (one from each class 4,3,2 respectively) which showcase the largest difference between the two algorithms used. Table \ref{tab:sim_data_res} presents the coordinates of the 3 points chosen and the arrows on figure \ref{fig:exp1} show the counterfactual point in the improved rating class. The purple arrow is the $\delta_x$ from the graduate descent, while the yellow arrow depicts the sparsity algorithm result. Table \ref{tab:sim_data_res} gives the numerical values of the $\delta_x$ and shows that the ratings are improved successfully. 

The gradient descent solution ``improves'' the class by changing all coordinates. The largest changes are in the first two coordinates, as they should, while the remaining three coordinates are just noise.  
Compared to the solution from the gradient descent, the sparsity algorithm solution removes the ``noise'' from features. It picks the relevant coordinate to be changed every time. 

However, we also showcase an exception (point 3). For this point the rating indeed improves from $2$ to $1$. However, the algorithm picks the feature $x_3$ to change in addition to $x_1$. This is due to the fact that the relative change ratio $\delta_x'$ is larger for $x_3$ than for $x_1$ just by chance. The sparsity algorithm checks the result for $x_3$ which does not change the rating, then at the next iteration it settles on the solution. 

This point 3 is one of few exceptions we observed in our results. It actually illustrates an issue we will observe in the next case study dealing with real data. 

% Table generated by Excel2LaTeX from sheet 'simulated_example'
\begin{table}[htbp]
  \centering
  \caption{Sample result for simulated data}
    \begin{tabular}{l|rrrrr|r}
    \toprule
          & \multicolumn{1}{l}{$x_1$} & \multicolumn{1}{l}{$x_2$} & \multicolumn{1}{l}{$x_3$} & \multicolumn{1}{l}{$x_4$} & \multicolumn{1}{l|}{$x_5$} & \multicolumn{1}{l}{Rating} \\
    \midrule
    original vector & 0.6019 & -0.4742 & 0.0827 & -0.0595 & 0.0588 & 4 \\
    GD solution ($\delta_x$) & -0.767 & -0.5539 & 0.0179 & -0.0095 & -0.012 & 3 \\
    Optimized Algo ($\delta_x$) & -0.767 & 0     & 0     & 0     & 0     & 3 \\
    \midrule
    original vector & -0.5488 & -1.0176 & 0.1723 & 0.2329 & 0.4329 & 3 \\
    GD solution ($\delta_x$) & -0.3963 & 1.0276 & -0.0133 & 0.0149 & 0.0074 & 2 \\
    Optimized Algo ($\delta_x$) & 0     & 1.0276 & 0     & 0     & 0     & 2 \\
    \midrule
    original vector & -1.3814 & 0.5363 & 0.0031 & -0.2783 & -0.074 & 2 \\
    GD solution ($\delta_x$) & 1.5275 & 0.4331 & 0.0054 & -0.0134 & -0.0071 & 1 \\
    Optimized Algo ($\delta_x$) & 1.5275 & 0     & 0.0054 & 0     & 0     & 1 \\
    \bottomrule
    \end{tabular}%
  \label{tab:sim_data_res}%
\end{table}%

This study primarily focuses on the L0 norm of $\delta_x$ as a measurement of the effort defined in section \ref{why_sig}. Recall that the objective of our problem in equation \eqref{eq: origin} is to increase the sparsity of the solution. However, the $L_2$ norm may be viewed as another measurement of effort as it calculates the total `distance' between the original point and the target. 

%We are not trying to minimize $L_2$ in our problem. 
%While L0 will be improved by design, $L_2$ may not.

To formally compare the solution from the gradient descent with the solution from the sparsity algorithm we perform matched pairs one-sided t tests as follows:

\paragraph{ L0 testing}
\begin{small}
\begin{align*}
   & H_0: \text{ The difference between L0 distance obtained from gradient descent method and L0 distance obtained from sparsity  } \\
   & \text{algorithm is equal to $0$}\\
    &H_a: \text{ The difference between L0 distance obtained from gradient descent method and L0 distance obtained from sparsity } \\
    & \text{algorithm is less than $0$}
\end{align*}
\end{small}

\paragraph{ L2 testing}
\begin{small}
\begin{align*}
   & H_0: \text{ The difference between $L_2$ distance obtained from gradient descent method and $L_2$ distance obtained from sparsity  } \\
   & \text{algorithm is equal to $0$}\\
    &H_a: \text{ The difference between $L_2$ distance obtained from gradient descent method and $L_2$ distance obtained from sparsity  } \\
    & \text{algorithm is greater than $0$}
\end{align*}
\end{small}

We use all the points in the dataset to perform these tests. We treat each change separately - from  $4$ to $3$, $3$ to $2$, and $2$ to $1$, respectively. The average L0 and $L_2$ for each group and the results for the matched pairs t-tests are presented in Table \ref{tab:sim_ttest}. 

From these results it is evident  that the solution from the sparsity algorithm is significantly smaller than the solution from the gradient descent, i.e., requires less ``effort''. 

% % Table generated by Excel2LaTeX from sheet 'exp1_res'
% \begin{table}[htbp]
%   \centering
%   \caption{Results of testing whether there is a difference in the procedures observed in L0 and the L2 norm}
%   \label{tab:sim_ttest}%
%   \begin{threeparttable}
%     \begin{tabular}{lrr|rr}
% \cmidrule{2-5}          & \multicolumn{1}{l}{L2 from sparsity} & \multicolumn{1}{l|}{L2 from GD} & \multicolumn{1}{l}{L0 from sparsity} & \multicolumn{1}{l}{L0 from GD} \\
%     \midrule
%     2 to 1  & 1.1518* & 1.1541 & 1.1526* & 5.000 \\
%     3 to 2  & 1.0764* & 1.0783 & 1.1667* & 5.000 \\
%     4 to 3  & 1.1961* & 1.2000 & 1.1839* & 5.000 \\
%     \bottomrule

%     \end{tabular}%

%       \begin{tablenotes}
%       \small
%       \item *significant at $p < 0.01$
%     \end{tablenotes}
%     \end{threeparttable}
  
% \end{table}%

% Table generated by Excel2LaTeX from sheet 'exp1_res'
\begin{table}[htbp]
  \centering
  \caption{Results of testing whether there is a difference in the procedures observed in L0 and the L2 norm}
    \begin{tabular}{lrrl|rrl}
\cmidrule{2-7}          & \multicolumn{1}{l}{L2 from } & \multicolumn{1}{l}{L2 from } & L2\_diff & \multicolumn{1}{l}{L0 from} & \multicolumn{1}{l}{L0 from } & L0\_diff \\
       & \multicolumn{1}{r}{sparsity} & \multicolumn{1}{r}{ GD} &  & \multicolumn{1}{l}{sparsity} & \multicolumn{1}{l}{GD} &  \\
    \midrule
    2 to 1 & 1.15182 & 1.15409 & 0.00227 (0.00024) & 1.15255 & 5.00000 & 3.84745 (0.00693) \\
    3 to 2 & 1.07638 & 1.07827 & 0.00189 (0.00022) & 1.16672 & 5.00000 & 3.83328 (0.00728) \\
    4 to 3 & 1.19610 & 1.20001 & 0.00390 (0.00040) & 1.18393 & 5.00000 & 3.81607 (0.00758) \\
    \bottomrule
    \end{tabular}%
  \label{tab:sim_ttest}%
\end{table}%

%The point of this table is to show that the sparsity algorithm produce significantly different changes to the input variables. 

% % Table generated by Excel2LaTeX from sheet 'Sheet1'
% \begin{table}[htbp]
%   \centering
%   \caption{One side t-test results}
%     \begin{tabular}{c|rr|rr|r}
% \cmidrule{2-6}          & \multicolumn{2}{c|}{Mean} & \multicolumn{2}{c|}{STD} & \multicolumn{1}{l}{p-value} \\
% \cmidrule{2-6}          & \multicolumn{1}{c}{ori} & \multicolumn{1}{c|}{dan} & \multicolumn{1}{c}{ori} & \multicolumn{1}{c|}{dan} &  \\
%     \midrule
%     2to1  & 1.2411 & 1.1891 & 0.3787 & 0.3765 & 4.39E-08 \\
%     3to2  & 1.1650 & 1.1161 & 0.3814 & 0.3768 & 3.14E-07 \\
%     4to3  & 1.3061 & 1.2469 & 0.4085 & 0.4038 & 6.09E-09 \\
%     \bottomrule
%     \end{tabular}%
%   \label{tab:sim_ttest}%
% \end{table}%

\subsection{Case study: Quarterly financial statement data}

\paragraph{A description of the Financial statement data used} \label{fin_data}

In this section we apply the sparsity algorithm to data obtained from financial statements. Given a particular financial statement, there may be many ways in which to improve the financial stability of a company and thus  increasing its credit rating. In this work, we are trying to provide a data driven answer which is based purely on the machine learning technique used. To this end, we have to assume that the machine learning technique used to determine the original $f$ is very accurate. Recall that the counterfactual problem we are focusing on, is defined for a given $f$. 

We apply the methodology to companies chosen from 3 sectors of the US economy: Financial, Healthcare, and Information Technology (IT). We first clean the data by removing features which are not reported for each of the specific sectors. The data is thus reduced to around 300 variables for each sector (294, 296, 296 respectively). Next, we define $w$ in equation \eqref{eq: mod1} by analyzing all remaining features for each sector and determining whether each feature can be feasibly changed. More precisely, certain accounting variables may not be changed because of contractual obligation, unpredictable events, related to tax, city governance, etc. Table \ref{tab:reasons} groups all the reasons we found as to why accounting variables may not be changed in practice. The table also lists one accounting variable as an example for each of the reasons. A complete list of accounting variables that we found hard or impossible to change is presented in Table \ref{tab:all_fea_not_manage} of the Appendix \ref{sec:appendix}. The remaining number of variables that are not masked by $w$ is $87$, $87$, and $86$ respectively for Healthcare, IT and Finance sectors.

\begin{table}[htbp]
  \centering
  \caption{The list of reasons why the respective variable may not be feasibly changed }
    \begin{tabular}{p{19.165em}l}
    \toprule
    \multicolumn{1}{l}{Reasons} & Example  \\
    \midrule
    \multicolumn{1}{l}{Scheduled items} & \textit{Pension Plan} \\
    \multicolumn{1}{l}{Assets are discontinued operations} & \textit{Extraordinary Items and Discontinued Operations} \\
    \multicolumn{1}{l}{Intangible asset} & \textit{Good Will} \\
    \multicolumn{1}{l}{Special items} & \textit{Costs of Failed Acquisitions} \\
    \multicolumn{1}{l}{Regulated items} & \textit{Tier 1 Capital Ratio} \\
    \multicolumn{1}{l}{Agreements with shareholders, employees} & \textit{Deferred Compensation} \\
    \multicolumn{1}{l}{Computational Items} & \textit{Depreciation \& Amortization} \\
    \multicolumn{1}{l}{Special events} & \textit{Loss from Flood/Fire} \\
    \multicolumn{1}{l}{Loss/gain from subsidiary} & \textit{Equity in Earnings (I/S) - Unconsolidated Subsidiaries} \\
    Non-operating items & \textit{Gain/Loss on Sale of Property} \\
    \bottomrule
    \end{tabular}%
  \label{tab:reasons}%
\end{table}%

\subsubsection{Question 2: Comparing the results of the algorithms with quarters when companies changed ratings.}\label{exp2:real_rating}

%This section investigates our algorithm by applying into financial statement data describe in \ref{fin_data}. Specifically, we focus on those companies whose credit ratings are improved at next quarter. We implement our algorithm to find the smallest effort on current manageable financial statement to ensure the counterfactual financial variables are qualified a better credit rating. 

% Table generated by Excel2LaTeX from sheet 'exp2'

It is simple to visualize the results of the algorithms for the synthetically generated data. For real data, the clusters are hard to visualize, but the algorithms works in a similar way. In most cases we are able to determine a $\delta_x$ which improves rating using either the gradient descent (GD) or the sparsity algorithm. However, how relevant is this $\delta_x$? Suppose we find a company to listen to our advice and for the next quarter the company places resources towards changing the variables indicated by $\delta_x$. If the targets are reached would the company improve its rating during the next quarter? 

This is of course a question hard to answer. In an attempt to answer it, we focus on those quarters and companies whose ratings actually improved during the next quarter. We apply GD and the sparsity algorithms to the statements from the quarters before the rating change. To assess the effectiveness of the proposed changes we calculate the actual \textbf{Real change} between the two consecutive quarters when the ratings improved. Table \ref{tab:exp2_ignore}  presents these values in columns 1 and 2. For example, the L0 number for the Healthcare sector Real change is calculated by looking at how many features changed between the two consecutive quarters when the rating of the company went up. We display the average number of features changed for all companies in the healthcare sector which went up in ratings.  We compare this real change ``effort'' with the proposed changes by the two algorithms in columns 3 and 4 of the table. The numbers in these columns are calculated using only the data from the quarters before the rating changed. Mathematically, they are calculated as the respective norm of $w\circ \delta_x$. Since both Gradient Descent (GD) and Sparsity algorithms only change a selected number of features (the unmasked features), for a proper comparison in table \ref{tab:exp2_ignore} we calculate the Real change only for the features that can be changed (column two).  Similar numbers are calculated for all sectors.

\begin{table}[htbp]
  \centering
  \caption{Results comparing with the real rating change when ignoring 0's in the feasible data }
    \begin{tabular}{cl|rrrrc}
\cmidrule{3-7}          & \multicolumn{1}{r}{} & \multicolumn{1}{p{5.165em}}{Real change} & \multicolumn{1}{p{7em}}{Real change for relevant variables} & \multicolumn{1}{p{7em}}{Change for GD} & \multicolumn{1}{p{7em}}{Change for Sparsity} & \multicolumn{1}{p{7em}}{Match Rate} \\
    \midrule
    \multirow{2}[2]{*}{Healthcare} & L0    & 113.84 & 59.02 & 87.00 & 53.82 & \multirow{2}[2]{*}{85.43\%} \\
          & L2    & 4744.44 & 4263.46 & 6021.42 & 4615.10 &  \\
    \midrule
    \multirow{2}[2]{*}{IT} & L0    & 119.13 & 61.95 & 87.00 & 60.12 & \multirow{2}[2]{*}{87.41\%} \\
          & L2    & 12550.07 & 11774.29 & 2727.35 & 2057.80 &  \\
    \midrule
    \multirow{2}[2]{*}{Financial} & L0    & 101.10 & 48.39 & 86.00 & 57.24 & \multirow{2}[2]{*}{76.58\%} \\
          & L2    & 65607.00 & 46018.43 & 11474.30 & 7591.71 &  \\
    \bottomrule
    \end{tabular}%
  \label{tab:exp2_ignore}%
\end{table}%

The last column in the table \ref{tab:exp2_ignore} is labeled Match Rate. For each company that changed rating we look at the features suggested to be changed by the sparsity algorithms. We calculate what percentage of them were actually changed in the real statements between the two quarters when ratings improved. A high match rate indicate that the features selected by the sparsity algorithm are similar to the changed features in the real statements. It is worth mentioning (again) that the sparsity algorithm comes up with these features based solely on the data from the quarter BEFORE the ratings changed. 

However, to obtain the numbers in the table \ref{tab:exp2_ignore} we implement the sparsity algorithm with a different Step 4 mentioned in remark \ref{step4}. Specifically, step 4 is: ``we set $\delta_{x_i}' = 0$ when $x_i = 0$''. With this change the sparsity algorithm essentially ignores those feasible features in the original statement $x$ whose value is $0$. We do this inspired by the synthetic data in the previous section. Recall the point 3 which by chance had the 3rd coordinate with a large relative change. A similar phenomenon is happening in the real statement data when there are $0$'s present in the unmasked set of features. The algorithm focuses on them as the relative change from $0$ is technically infinite. In fact, in the real statements some of those $0$ features do change, and that is probably why we aren't able to capture 100\% of the changes. 

Looking at L0 (the number of changed features) the results are consistent. In the two quarters data the average number of features changed is between $100$ and $120$, while the number of relevant features changed is about half the total number. The gradient descent changes all features and in fact only the $L_2$ distance is relevant for it. However, the sparsity algorithm produces a number of features to be changed which is similar to the real number. Although it is nice to see that we recover most of the features that actually changed this is not our goal, as we want to identify the smallest number of changes possible with a minimum effort. By neglecting the features with a $0$ value, the algorithm is probably ignoring features that might be very important to improve credit rating. This is why we are replacing the step 4 in the sparsity algorithm with ``We set $\delta_{x_i}'= 1$ when the component $x_i= 0$'', as it was in fact written in the actual algorithm. We rerun this algorithm and present the results in Table \ref{tab:exp2}. 

\begin{table}[htbp]
  \centering
  \caption{Results comparing with the real rating change}
    \begin{tabular}{cc|ccccc}
\cmidrule{3-7}          & \multicolumn{1}{r}{} & \multicolumn{1}{p{5.165em}}{Real change} & \multicolumn{1}{p{7em}}{Real change for relevant variables} & \multicolumn{1}{p{7em}}{Change for GD} & \multicolumn{1}{p{7em}}{Change for Sparsity} & \multicolumn{1}{p{7em}}{Match Rate} \\
    \midrule
    \multirow{2}[2]{*}{Healthcare} & L0    & 113.84 & 59.02 & 87.00 & 22.93 & \multirow{2}[2]{*}{61.01\%}\\
          & L2    & 4744.44 & 4263.46 & 6021.42 & 5239.12 &  \\
    \midrule
    \multirow{2}[2]{*}{IT} & L0    & 119.13 & 61.95 & 87.00 & 24.73 & \multirow{2}[2]{*}{52.59\%} \\
          & L2    & 12550.07 & 11774.29 & 2727.35 & 1925.80 &  \\
    \midrule
    \multirow{2}[2]{*}{Financial} & L0    & 101.10 & 48.39 & 86.00 & 33.46 & \multirow{2}[2]{*}{37.35\%} \\
          & L2    & 65607.00 & 46018.43 & 11474.30 & 7962.78 &  \\
    \bottomrule
    \end{tabular}%
  \label{tab:exp2}%
\end{table}%

Including the $0$ features in the set of possible changes helps the sparsity algorithm reduce its L0 norm. However, the match rate of the sparsity algorithm drops to around $50\%$. Considering the purpose of the algorithm is to identify relevant features for improving ratings the sparsity of the $\delta_x$ is important. In terms of the $L_2$ norm, we note that the magnitude of change (``effort'') is reduced dramatically in the Finance and IT sectors but it is in fact increased on the average in the Healthcare sector. We believe this it normal as focusing on improving the sparsity of the solution, the feasible domain would be reduced. Thus, to qualify the solution for an improved rating, we have to exert more effort on those feasible features. This may cause an increase in the $L_2$ norm of the solution.

Comparatively, we observe that the companies in the financial sector need to exert more effort to improve their credit rating than companies in the IT and healthcare sectors.

\paragraph{Why two sparsity algorithms?} Generally, published articles do not detail all attempts and only showcase the best, which is typically the last algorithm. In this article, we chose to present a variant of the algorithm which we initially employed as well as the final algorithm version. We decided to do this as we are dealing with real data between two quarters, and our algorithm is dependent on how well the MLP $f$ is performing. Thus it is important to validate the features we obtain from using the algorithm on the previous quarter with the features actually changed in the next quarter. This match is an argument that our algorithm catches the relevant changes as well as that  $f(\cdot)$ is producing accurate results. 

This is also important for the final algorithm results in Table \ref{tab:exp2}. Taken by itself the sparsity algorithm results are academic. However, when corroborated with the results in Table \ref{tab:exp2_ignore} which show it is possible to match the real changes, we think the results point to a valid way to potentially produce an improved rating during the next quarter.

\subsubsection{Comparing the effort needed to improve ratings from different levels} 

In this section, we implement the sparsity algorithm to all observations in the dataset. We calculate the ``effort'' needed for a company during a particular quarter to improve during the next quarter. We aggregate the results by the specific rating and sector. We want to investigate how the effort changes depending on the ratings the company is at the time of the respective quarter. For example, is it harder to improve rating from the highest non investment grade  (BB+) to an investment grade (BBB-) then it is from other ratings?

%Table generated by Excel2LaTeX from sheet 'Sheet1'
\begin{table}[htbp]
  \centering
  \caption{S\&P rating description}
    \begin{tabular}{|c|c|c|}
    \toprule
    S\&P rating & \multicolumn{2}{c|}{Rating description} \\
    \midrule
    AAA   & Extremely strong capacity to meet its financial commitments & \multicolumn{1}{c|}{\multirow{10}[20]{*}{Investment grade}} \\
\cmidrule{1-2}    AA+   & \multicolumn{1}{c|}{\multirow{3}[6]{*}{Very strong capacity to meet its financial commitments}} &  \\
\cmidrule{1-1}    AA    &       &  \\
\cmidrule{1-1}    AA-   &       &  \\
\cmidrule{1-2}    A+    & \multicolumn{1}{c|}{\multirow{3}[6]{*}{Strong capacity to meet its financial commitments}} &  \\
\cmidrule{1-1}    A     &       &  \\
\cmidrule{1-1}    A-    &       &  \\
\cmidrule{1-2}    BBB+  & \multicolumn{1}{c|}{\multirow{3}[6]{*}{Adequate capacity to meet its financial commitments}} &  \\
\cmidrule{1-1}    BBB   &       &  \\
\cmidrule{1-1}    BBB-  &       &  \\
    \midrule
    BB+   & \multicolumn{1}{c|}{\multirow{3}[6]{*}{ Has inadequate capacity to meet its financial commitments}} & \multicolumn{1}{c|}{\multirow{12}[24]{*}{Non investment grade }} \\
\cmidrule{1-1}    BB    &       &  \\
\cmidrule{1-1}    BB-   &       &  \\
\cmidrule{1-2}    B+    & \multicolumn{1}{c|}{\multirow{3}[6]{*}{Has the capacity to meet its financial commitments}} &  \\
\cmidrule{1-1}    B     &       &  \\
\cmidrule{1-1}    B-    &       &  \\
\cmidrule{1-2}    CCC+  & \multicolumn{1}{c|}{Substantial risks} &  \\
\cmidrule{1-2}    CCC   & \multicolumn{1}{c|}{Extremely speculative} &  \\
\cmidrule{1-2}    CCC-  & \multicolumn{1}{c|}{\multirow{3}[6]{*}{Default imminent with little prospect for recovery}} &  \\
\cmidrule{1-1}    CC    &       &  \\
\cmidrule{1-1}    C     &       &  \\
\cmidrule{1-2}    D     & In default &  \\
    \bottomrule
    \end{tabular}%
  \label{exp3:rating_desc}%
\end{table}%

For reference in Table \ref{exp3:rating_desc} we present the Standard \& Poor's classification and ratings interpretation. The higher the rating, the lower the interest rate the company has to pay. Furthermore, having an investment grade rating means that the pool of investors is enlarged considerably as government regulations prevent pension funds and mutual funds from purchasing non-investment grade bonds. In fact, if any of their holdings drops below BBB-, the pension funds are required to sell, often at a loss. 

\begin{table}[htbp]
  \centering
  \caption{The average ``effort'' required to improve ratings for each rating level}
    \begin{tabular}{cc|rrr|rrr|rrr}
\cmidrule{3-11}          &       & \multicolumn{3}{c|}{IT} & \multicolumn{3}{c|}{Healthcare} & \multicolumn{3}{c}{Financial} \\
    \midrule
    ori   & curr  & \multicolumn{1}{c}{L2} & \multicolumn{1}{c}{L0} & \multicolumn{1}{c|}{Count} & \multicolumn{1}{c}{L2} & \multicolumn{1}{c}{L0} & \multicolumn{1}{c|}{Count} & \multicolumn{1}{c}{L2} & \multicolumn{1}{c}{L0} & \multicolumn{1}{c}{Count} \\
    \midrule
    \multicolumn{1}{l}{AA+} & \multicolumn{1}{l|}{AAA} & 108400.3 & 27.3  & 15    &       &       &       & 11183.0 & 30.1  & 13 \\
    \multicolumn{1}{l}{AA} & \multicolumn{1}{l|}{AA+} & 94786.6 & 52.0  & 47    & 174127.2 & 62.7  & 128   & 41469.8 & 34.7  & 85 \\
    \multicolumn{1}{l}{AA-} & \multicolumn{1}{l|}{AA} & 59565.2 & 38.5  & 59    & 4879.0 & 37.2  & 126   & 37547.4 & 43.2  & 382 \\
    \multicolumn{1}{l}{A+} & \multicolumn{1}{l|}{AA-} & 23463.1 & 41.8  & 297   & 4854.6 & 28.1  & 268   & 6051.8 & 35.8  & 641 \\
    \multicolumn{1}{l}{A} & \multicolumn{1}{l|}{A+} & 3482.2 & 31.1  & 194   & 2668.8 & 21.2  & 351   & 15756.5 & 31.1  & 630 \\
    \multicolumn{1}{l}{A-} & \multicolumn{1}{l|}{A} & 1372.4 & 28.1  & 217   & 2892.9 & 29.6  & 192   & 6755.0 & 29.1  & 644 \\
    \multicolumn{1}{l}{BBB+} & \multicolumn{1}{l|}{A-} & 2025.7 & 32.4  & 234   & 1096.2 & 25.8  & 308   & 4483.3 & 31.0  & 421 \\
    \multicolumn{1}{l}{BBB} & \multicolumn{1}{l|}{BBB+} & 1250.3 & 28.1  & 342   & 1598.6 & 24.5  & 343   & 2492.2 & 31.8  & 353 \\
    \multicolumn{1}{l}{BBB-} & \multicolumn{1}{l|}{BBB} & 386.7 & 22.1  & 195   & 1169.3 & 21.9  & 244   & 2886.5 & 27.8  & 212 \\
    \multicolumn{1}{l}{BB+} & \multicolumn{1}{l|}{BBB-} & 1474.1 & 32.9  & 180   & 602.3 & 23.8  & 125   & 1637.3 & 36.5  & 80 \\
    \multicolumn{1}{l}{BB} & \multicolumn{1}{l|}{BB+} & 864.0 & 25.9  & 74    & 567.5 & 23.6  & 182   & 6191.6 & 47.0  & 19 \\
    \multicolumn{1}{l}{BB-} & \multicolumn{1}{l|}{BB} & 1121.8 & 29.0  & 146   & 262.6 & 16.4  & 90    & 4243.2 & 47.7  & 11 \\
    \multicolumn{1}{l}{B+} & \multicolumn{1}{l|}{BB-} & 411.0 & 28.9  & 112   & 1344.1 & 20.5  & 45    & 1943.0 & 43.1  & 20 \\
    \multicolumn{1}{l}{B} & \multicolumn{1}{l|}{B+} & 402.6 & 16.6  & 39    & 517.9 & 46.7  & 15    & 4176.9 & 63.5  & 2 \\
    \multicolumn{1}{l}{B-} & \multicolumn{1}{l|}{B} & 950.7 & 33.9  & 24    &       &       &       & 1817.3 & 34.5  & 15 \\
    \multicolumn{1}{l}{CCC+} & \multicolumn{1}{l|}{B-} & 1033.3 & 36.0  & 12    &       &       &       & 1033.4 & 41.1  & 9 \\
    \midrule
    \multicolumn{2}{c|}{Average} & 8735.4 & 31.3  &       & 11228.0 & 27.1  &       & 11288.6 & 33.2  &  \\
    \bottomrule
    \end{tabular}%
  \label{tab:exp3_first}%
\end{table}%

Table \ref{tab:exp3_first} presents the L0 and $L_2$ averages for the sparsity algorithm as well as how many observations were in each category (``Count'' column). We observe no clear pattern to indicate that the results fit with the ratings in table \ref{exp3:rating_desc}. We generally note that the ``effort'' needed increases when ratings are increasing. This suggests that it may be easier to increase in rating from say BB+ to BBB- than it is to increase from A+ to AA-.

However, we need to point out an issue that arises when we aggregate all these companies. Specifically, a particular credit rating says that a company is in a range of risk levels, it is not providing a specific value of risk for that company. Thus, the actual risk value for companies within the same credit rating score may be different. For example, a company XXX may be at one extreme of ranges for the AA ratings, while company YYY may be at the other extreme and still rated AA. It is obviously more difficult for one company to improve its ratings than it is for the other company. The results in the Table \ref{tab:exp3_first} contain all the companies for a particular rating range and it calculates an aggregate average effort. It may be impossible or very hard for a company that just improved its rating to go to an even better rating the following quarter. In the sparsity algorithm, $\lambda$ in equation \eqref{eq: mod1} controls this ``difficulty''. A higher $\lambda$ indicates this particular company is harder to improve its rating level. Thus, the  ``effort'' needed may be larger.
\begin{table}[!htbp]
  \centering
  \caption{The number of companies improving rating as $\lambda$ changes in the IT sector. }
    \begin{tabular}{c|rrrrrrrrrrrrrrrr}
    \toprule
    Lambda & \multicolumn{1}{c}{AAA} & \multicolumn{1}{c}{AA+} & \multicolumn{1}{c}{AA} & \multicolumn{1}{c}{AA-} & \multicolumn{1}{c}{A+} & \multicolumn{1}{c}{A} & \multicolumn{1}{c}{A-} & \multicolumn{1}{c}{BBB+} & \multicolumn{1}{c}{BBB} & \multicolumn{1}{c}{BBB-} & \multicolumn{1}{c}{BB+} & \multicolumn{1}{c}{BB} & \multicolumn{1}{c}{BB-} & \multicolumn{1}{c}{B+} & \multicolumn{1}{c}{B} & \multicolumn{1}{c}{B-} \\
    \midrule
    0.1   & 0     & 0     & 0     & 0     & 0     & 0     & 0     & 0     & 1     & 0     & 0     & 0     & 0     & 0     & 0     & 0 \\
    5     & 0     & 0     & 0     & 0     & 0     & 0     & 0     & 0     & 0     & 0     & 8     & 0     & 0     & 0     & 0     & 0 \\
    10    & 0     & 0     & 0     & 0     & 2     & 6     & 2     & 0     & 0     & 0     & 7     & 4     & 0     & 18    & 0     & 0 \\
    50    & 0     & 0     & 1     & 2     & 49    & 159   & 135   & 122   & 105   & 69    & 42    & 72    & 85    & 21    & 22    & 9 \\
    100   & 0     & 0     & 0     & 22    & 55    & 47    & 46    & 114   & 81    & 63    & 10    & 51    & 19    & 0     & 1     & 1 \\
    200   & 0     & 0     & 11    & 65    & 46    & 5     & 29    & 80    & 7     & 29    & 6     & 18    & 8     & 0     & 1     & 2 \\
    500   & 0     & 0     & 5     & 95    & 17    & 0     & 17    & 24    & 1     & 11    & 1     & 1     & 0     & 0     & 0     & 0 \\
    1000  & 0     & 5     & 6     & 69    & 22    & 0     & 5     & 2     & 0     & 7     & 0     & 0     & 0     & 0     & 0     & 0 \\
    10000 & 15    & 41    & 36    & 43    & 3     & 0     & 0     & 0     & 0     & 1     & 0     & 0     & 0     & 0     & 0     & 0 \\
    100000 & 0     & 1     & 0     & 1     & 0     & 0     & 0     & 0     & 0     & 0     & 0     & 0     & 0     & 0     & 0     & 0\\
    \bottomrule
    \end{tabular}%
  \label{tab:labda_it}%
  \begin{tablenotes}
      \footnotesize
      \item \textit{Higher lambda means more effort needs to be exerted to improve rating.}
    \end{tablenotes}
\end{table}%

\begin{table}[!htbp]
  \centering
  \caption{The number of companies improving rating as $\lambda$ changes in the Financial sector. }
    \begin{tabular}{c|rrrrrrrrrrrrrrrr}
    \toprule
    Lambda & \multicolumn{1}{c}{AAA} & \multicolumn{1}{c}{AA+} & \multicolumn{1}{c}{AA} & \multicolumn{1}{c}{AA-} & \multicolumn{1}{c}{A+} & \multicolumn{1}{c}{A} & \multicolumn{1}{c}{A-} & \multicolumn{1}{c}{BBB+} & \multicolumn{1}{c}{BBB} & \multicolumn{1}{c}{BBB-} & \multicolumn{1}{c}{BB+} & \multicolumn{1}{c}{BB} & \multicolumn{1}{c}{BB-} & \multicolumn{1}{c}{B+} & \multicolumn{1}{c}{B} & \multicolumn{1}{c}{B-} \\
    \midrule
    10    & 0     & 0     & 0     & 0     & 1     & 0     & 4     & 0     & 0     & 0     & 0     & 0     & 0     & 0     & 0     & 0 \\
    50    & 0     & 1     & 2     & 54    & 82    & 85    & 94    & 63    & 95    & 6     & 5     & 0     & 2     & 0     & 3     & 2 \\
    100   & 0     & 7     & 10    & 106   & 123   & 82    & 48    & 82    & 79    & 5     & 1     & 0     & 4     & 0     & 0     & 3 \\
    200   & 1     & 1     & 52    & 118   & 140   & 127   & 82    & 89    & 13    & 17    & 1     & 0     & 9     & 1     & 8     & 4 \\
    500   & 9     & 6     & 108   & 183   & 135   & 125   & 99    & 76    & 25    & 43    & 9     & 8     & 4     & 1     & 4     & 0 \\
    1000  & 3     & 3     & 61    & 84    & 37    & 104   & 42    & 27    & 0     & 9     & 3     & 3     & 1     & 0     & 0     & 0 \\
    10000 & 0     & 40    & 122   & 87    & 72    & 90    & 42    & 15    & 0     & 0     & 0     & 0     & 0     & 0     & 0     & 0 \\
    100000 & 0     & 27    & 27    & 9     & 40    & 31    & 10    & 1     & 0     & 0     & 0     & 0     & 0     & 0     & 0     & 0 \\
    \bottomrule
    \end{tabular}%
  \label{tab:labda_fin}%
  \begin{tablenotes}
      \small
      \item \textit{Higher lambda means more effort needs to be exerted to improve rating.}
    \end{tablenotes}
\end{table}%

\begin{table}[ht]
  \centering
  \caption{The number of companies improving rating as $\lambda$ changes in the Healthcare sector. }
    \begin{tabular}{c|rrrrrrrrrrrrr}
    \toprule
    Lambda & \multicolumn{1}{c}{AA+} & \multicolumn{1}{c}{AA} & \multicolumn{1}{c}{AA-} & \multicolumn{1}{c}{A+} & \multicolumn{1}{c}{A} & \multicolumn{1}{c}{A-} & \multicolumn{1}{c}{BBB+} & \multicolumn{1}{c}{BBB} & \multicolumn{1}{c}{BBB-} & \multicolumn{1}{c}{BB+} & \multicolumn{1}{c}{BB} & \multicolumn{1}{c}{BB-} & \multicolumn{1}{c}{B+} \\
    \midrule
    5     & 0     & 0     & 0     & 0     & 0     & 0     & 0     & 0     & 0     & 1     & 1     & 0     & 0 \\
    10    & 0     & 1     & 0     & 0     & 0     & 0     & 0     & 0     & 0     & 0     & 0     & 0     & 0 \\
    50    & 0     & 7     & 36    & 18    & 11    & 63    & 45    & 79    & 23    & 34    & 23    & 20    & 11 \\
    100   & 0     & 6     & 50    & 78    & 32    & 107   & 53    & 74    & 54    & 57    & 46    & 16    & 3 \\
    200   & 0     & 26    & 73    & 63    & 62    & 66    & 83    & 43    & 36    & 68    & 20    & 9     & 1 \\
    500   & 0     & 49    & 62    & 100   & 45    & 55    & 110   & 47    & 12    & 22    & 0     & 0     & 0 \\
    1000  & 1     & 35    & 29    & 44    & 34    & 14    & 34    & 1     & 0     & 0     & 0     & 0     & 0 \\
    10000 & 63    & 2     & 18    & 47    & 8     & 3     & 14    & 0     & 0     & 0     & 0     & 0     & 0 \\
    100000 & 64    & 0     & 0     & 1     & 0     & 0     & 4     & 0     & 0     & 0     & 0     & 0     & 0 \\
    \bottomrule
    \end{tabular}%
  \label{tab:labda_hea}%
  \begin{tablenotes}
      \small
      \item \textit{Higher lambda means more effort needs to be exerted to improve rating.}
    \end{tablenotes}
\end{table}%

Tables \ref{tab:labda_it}, \ref{tab:labda_fin}, and \ref{tab:labda_hea} present in each row the numbers of companies that successfully improved their current rating, for a particular $\lambda$ value. The tables are split by sector. We can see that as the $\lambda$ values increase more companies improve their rating. Recall our range of rating scores assertion. We interpret the values in the tables as the companies that are closer to the threshold (smaller lambda) are improving easier and thus are in an upper row in the tables. 

Looking at all the three tables we see the numbers shifting to left as $\lambda$ increases. This is consistent with our previous observations in table \ref{tab:exp3_first}. Indeed, the results seem to indicate that a lower $\lambda$ is needed (thus a lower effort) for the majority of the lower rated companies. In contrast a larger $\lambda$ value is needed for the majority of the high rated companies to improve their score. Thus, as the rating of the company gets better, a much larger effort is needed to further improve its ratings.

\section{Conclusion}

In this work we propose a sparsity algorithm that finds a counterfactual explanation for the credit rating problem. The sparsity algorithm is designed to discover the least amount of changes to be made to a particular financial statement variables that has a large probability of improving the prediction to a predefined credit rating. 

We apply the sparsity algorithm to a synthetically generated dataset as well as to quarterly financial statements data. Our toy case study, using synthetically generated data, shows that the sparsity algorithm can successfully change points to the target class, with less ``effort'' then the solution obtained using a gradient descent method. The results obtained using  quarterly financial statements confirm that the sparsity algorithm may be employed to significantly reduce the ``effort'' to improve corporate credit rating. More importantly, when analyzing quarterly statements before an actual rating increase we show that the sparsity algorithm captures the majority of features that in fact will have changed in the next quarter statement. This result gives us confidence to propose the final algorithm which results in an even more focused recommendation to the corporation's managers. 

Finally, we find that the ``effort'' required to improve the credit rating is positively related to the credit rating level. Specifically, improving credit rating for A rated corporations is much harder than improving credit rating for B level companies.

\section*{Acknowledgment}

The authors would like to acknowledge the UBS research grant awarded to the Hanlon Laboratories which provided partial support for this research. We want to acknowledge Bingyang Wen who provided helpful discussions about the algorithm. We also acknowledge Professor Zachary Feinstein who suggested the use of the $L_1$ norm in the proposed optimization problem.

\newpage

\bibliographystyle{chicago}
\bibliography{main}

\begin{thebibliography}{}

\bibitem[\protect\citeauthoryear{Addo, Guegan, and Hassani}{Addo
  et~al.}{2018}]{addo2018credit}
Addo, P.~M., D.~Guegan, and B.~Hassani (2018).
\newblock Credit risk analysis using machine and deep learning models.
\newblock {\em Risks\/}~{\em 6\/}(2), 38.

\bibitem[\protect\citeauthoryear{Ahn and Kim}{Ahn and
  Kim}{2011}]{ahn2011corporate}
Ahn, H. and K.-J. Kim (2011).
\newblock Corporate credit rating using multiclass classification models with
  order information.
\newblock {\em World Academy of Science, Engineering and Technology,
  International Journal of Social, Behavioral, Educational, Economic, Business
  and Industrial Engineering\/}~{\em 5\/}(12), 1783--1788.

\bibitem[\protect\citeauthoryear{Akdemir and Karsl{\i}}{Akdemir and
  Karsl{\i}}{2012}]{akdemir2012assessment}
Akdemir, A. and D.~Karsl{\i} (2012).
\newblock An assessment of strategic importance of credit rating agencies for
  companies and organizations.
\newblock {\em Procedia-Social and Behavioral Sciences\/}~{\em 58}, 1628--1639.

\bibitem[\protect\citeauthoryear{Bruckstein, Donoho, and Elad}{Bruckstein
  et~al.}{2009}]{bruckstein2009sparse}
Bruckstein, A.~M., D.~L. Donoho, and M.~Elad (2009).
\newblock From sparse solutions of systems of equations to sparse modeling of
  signals and images.
\newblock {\em SIAM review\/}~{\em 51\/}(1), 34--81.

\bibitem[\protect\citeauthoryear{Cai, Nie, and Huang}{Cai
  et~al.}{2013}]{cai2013exact}
Cai, X., F.~Nie, and H.~Huang (2013).
\newblock Exact top-k feature selection via l2, 0-norm constraint.
\newblock In {\em Twenty-third international joint conference on artificial
  intelligence}.

\bibitem[\protect\citeauthoryear{Carvalho, Pereira, and Cardoso}{Carvalho
  et~al.}{2019}]{carvalho2019machine}
Carvalho, D.~V., E.~M. Pereira, and J.~S. Cardoso (2019).
\newblock Machine learning interpretability: A survey on methods and metrics.
\newblock {\em Electronics\/}~{\em 8\/}(8), 832.

\bibitem[\protect\citeauthoryear{Cauchy et~al.}{Cauchy
  et~al.}{1847}]{cauchy1847methode}
Cauchy, A. et~al. (1847).
\newblock M{\'e}thode g{\'e}n{\'e}rale pour la r{\'e}solution des systemes
  d’{\'e}quations simultan{\'e}es.
\newblock {\em Comp. Rend. Sci. Paris\/}~{\em 25\/}(1847), 536--538.

\bibitem[\protect\citeauthoryear{Chakraborty, Tomsett, Raghavendra, Harborne,
  Alzantot, Cerutti, Srivastava, Preece, Julier, Rao, et~al.}{Chakraborty
  et~al.}{2017}]{chakraborty2017interpretability}
Chakraborty, S., R.~Tomsett, R.~Raghavendra, D.~Harborne, M.~Alzantot,
  F.~Cerutti, M.~Srivastava, A.~Preece, S.~Julier, R.~M. Rao, et~al. (2017).
\newblock Interpretability of deep learning models: a survey of results.
\newblock In {\em 2017 IEEE smartworld, ubiquitous intelligence \& computing,
  advanced \& trusted computed, scalable computing \& communications, cloud \&
  big data computing, Internet of people and smart city innovation
  (smartworld/SCALCOM/UIC/ATC/CBDcom/IOP/SCI)}, pp.\  1--6. IEEE.

\bibitem[\protect\citeauthoryear{Compustat}{Compustat}{2019}]{SP2019compustat}
Compustat, S. .~P. (2019).
\newblock {\em Compustat Online Manual}.
\newblock Standard \& Poor's.

\bibitem[\protect\citeauthoryear{Curry}{Curry}{1944}]{curry1944method}
Curry, H.~B. (1944).
\newblock The method of steepest descent for non-linear minimization problems.
\newblock {\em Quarterly of Applied Mathematics\/}~{\em 2\/}(3), 258--261.

\bibitem[\protect\citeauthoryear{Dittrich}{Dittrich}{2007}]{dittrich2007credit}
Dittrich, F. (2007).
\newblock {\em The credit rating industry: competition and regulation}.
\newblock Ph.\ D. thesis, Universit{\"a}t zu K{\"o}ln.

\bibitem[\protect\citeauthoryear{{General Data Protection Regulation}}{{General
  Data Protection Regulation}}{2016}]{regulation2016regulation}
{General Data Protection Regulation} (2016).
\newblock Regulation eu 2016/679 of the european parliament and of the council
  of 27 april 2016.
\newblock {\em Official Journal of the European Union. Available at: http://ec.
  europa. eu/justice/data-protection/reform/files/regulation\_oj\_en. pdf
  (accessed 20 September 2017)\/}.

\bibitem[\protect\citeauthoryear{Golbayani, Florescu, and Chatterjee}{Golbayani
  et~al.}{2020}]{golbayani2020comparative}
Golbayani, P., I.~Florescu, and R.~Chatterjee (2020).
\newblock A comparative study of forecasting corporate credit ratings using
  neural networks, support vector machines, and decision trees.
\newblock {\em The North American Journal of Economics and Finance\/}~{\em 54},
  101251.

\bibitem[\protect\citeauthoryear{Golbayani, Wang, and Florescu}{Golbayani
  et~al.}{2020}]{golbayani2020application}
Golbayani, P., D.~Wang, and I.~Florescu (2020).
\newblock Application of deep neural networks to assess corporate credit
  rating.
\newblock {\em arXiv preprint arXiv:2003.02334\/}.

\bibitem[\protect\citeauthoryear{Goodman and Flaxman}{Goodman and
  Flaxman}{2017}]{goodman2017european}
Goodman, B. and S.~Flaxman (2017).
\newblock European union regulations on algorithmic decision-making and a
  “right to explanation”.
\newblock {\em AI magazine\/}~{\em 38\/}(3), 50--57.

\bibitem[\protect\citeauthoryear{Goyal, Wu, Ernst, Batra, Parikh, and
  Lee}{Goyal et~al.}{2019}]{goyal2019counterfactual}
Goyal, Y., Z.~Wu, J.~Ernst, D.~Batra, D.~Parikh, and S.~Lee (2019).
\newblock Counterfactual visual explanations.
\newblock In {\em International Conference on Machine Learning}, pp.\
  2376--2384. PMLR.

\bibitem[\protect\citeauthoryear{Grath, Costabello, Van, Sweeney, Kamiab, Shen,
  and Lecue}{Grath et~al.}{2018}]{grath2018interpretable}
Grath, R.~M., L.~Costabello, C.~L. Van, P.~Sweeney, F.~Kamiab, Z.~Shen, and
  F.~Lecue (2018).
\newblock Interpretable credit application predictions with counterfactual
  explanations.
\newblock {\em arXiv preprint arXiv:1811.05245\/}.

\bibitem[\protect\citeauthoryear{H{\'a}jek and Olej}{H{\'a}jek and
  Olej}{2011}]{hajek2011credit}
H{\'a}jek, P. and V.~Olej (2011).
\newblock Credit rating modelling by kernel-based approaches with supervised
  and semi-supervised learning.
\newblock {\em Neural Computing and Applications\/}~{\em 20\/}(6), 761--773.

\bibitem[\protect\citeauthoryear{H{\'a}jek and Olej}{H{\'a}jek and
  Olej}{2014}]{hajek2014predicting}
H{\'a}jek, P. and V.~Olej (2014).
\newblock Predicting firms’ credit ratings using ensembles of artificial
  immune systems and machine learning--an over-sampling approach.
\newblock In {\em IFIP International Conference on Artificial Intelligence
  Applications and Innovations}, pp.\  29--38. Springer.

\bibitem[\protect\citeauthoryear{Huang, Zhang, Jiang, Stanforth, Welbl, Rae,
  Maini, Yogatama, and Kohli}{Huang et~al.}{2019}]{huang2019reducing}
Huang, P.-S., H.~Zhang, R.~Jiang, R.~Stanforth, J.~Welbl, J.~Rae, V.~Maini,
  D.~Yogatama, and P.~Kohli (2019).
\newblock Reducing sentiment bias in language models via counterfactual
  evaluation.
\newblock {\em arXiv preprint arXiv:1911.03064\/}.

\bibitem[\protect\citeauthoryear{Huang, Chen, Hsu, Chen, and Wu}{Huang
  et~al.}{2004}]{huang2004credit}
Huang, Z., H.~Chen, C.-J. Hsu, W.-H. Chen, and S.~Wu (2004).
\newblock Credit rating analysis with support vector machines and neural
  networks: a market comparative study.
\newblock {\em Decision support systems\/}~{\em 37\/}(4), 543--558.

\bibitem[\protect\citeauthoryear{Janocha and Czarnecki}{Janocha and
  Czarnecki}{2017}]{janocha2017loss}
Janocha, K. and W.~M. Czarnecki (2017).
\newblock On loss functions for deep neural networks in classification.
\newblock {\em arXiv preprint arXiv:1702.05659\/}.

\bibitem[\protect\citeauthoryear{Khashman}{Khashman}{2010}]{khashman2010neural}
Khashman, A. (2010).
\newblock Neural networks for credit risk evaluation: Investigation of
  different neural models and learning schemes.
\newblock {\em Expert Systems with Applications\/}~{\em 37\/}(9), 6233--6239.

\bibitem[\protect\citeauthoryear{Khemakhem and Boujelbene}{Khemakhem and
  Boujelbene}{2015}]{khemakhem2015credit}
Khemakhem, S. and Y.~Boujelbene (2015).
\newblock Credit risk prediction: A comparative study between discriminant
  analysis and the neural network approach.
\newblock {\em Accounting and Management Information Systems\/}~{\em 14\/}(1),
  60.

\bibitem[\protect\citeauthoryear{Kim and Sohn}{Kim and
  Sohn}{2010}]{kim2010support}
Kim, H.~S. and S.~Y. Sohn (2010).
\newblock Support vector machines for default prediction of smes based on
  technology credit.
\newblock {\em European Journal of Operational Research\/}~{\em 201\/}(3),
  838--846.

\bibitem[\protect\citeauthoryear{Kline and Berardi}{Kline and
  Berardi}{2005}]{kline2005revisiting}
Kline, D.~M. and V.~L. Berardi (2005).
\newblock Revisiting squared-error and cross-entropy functions for training
  neural network classifiers.
\newblock {\em Neural Computing \& Applications\/}~{\em 14\/}(4), 310--318.

\bibitem[\protect\citeauthoryear{Kumar and Bhattacharya}{Kumar and
  Bhattacharya}{2006}]{kumar2006artificial}
Kumar, K. and S.~Bhattacharya (2006).
\newblock Artificial neural network vs linear discriminant analysis in credit
  ratings forecast: A comparative study of prediction performances.
\newblock {\em Review of Accounting and Finance\/}~{\em 5\/}(3), 216--227.

\bibitem[\protect\citeauthoryear{Kumar and Haynes}{Kumar and
  Haynes}{2003}]{kumar2003forecasting}
Kumar, K. and J.~D. Haynes (2003).
\newblock Forecasting credit ratings using an ann and statistical techniques.
\newblock {\em International journal of business studies\/}~{\em 11\/}(1).

\bibitem[\protect\citeauthoryear{LeCun, Bengio, and Hinton}{LeCun
  et~al.}{2015}]{lecun2015deep}
LeCun, Y., Y.~Bengio, and G.~Hinton (2015).
\newblock Deep learning. nature 521 (7553), 436-444.
\newblock {\em Google Scholar Google Scholar Cross Ref Cross Ref\/}.

\bibitem[\protect\citeauthoryear{Luo and Chen}{Luo and
  Chen}{2019}]{luo2019bond}
Luo, H. and L.~Chen (2019).
\newblock Bond yield and credit rating: evidence of chinese local government
  financing vehicles.
\newblock {\em Review of Quantitative Finance and Accounting\/}~{\em 52\/}(3),
  737--758.

\bibitem[\protect\citeauthoryear{Mordvintsev, Olah, and Tyka}{Mordvintsev
  et~al.}{2015}]{mordvintsev2015inceptionism}
Mordvintsev, A., C.~Olah, and M.~Tyka (2015).
\newblock Inceptionism: Going deeper into neural networks.

\bibitem[\protect\citeauthoryear{Murdoch, Singh, Kumbier, Abbasi-Asl, and
  Yu}{Murdoch et~al.}{2019}]{murdoch2019interpretable}
Murdoch, W.~J., C.~Singh, K.~Kumbier, R.~Abbasi-Asl, and B.~Yu (2019).
\newblock Interpretable machine learning: definitions, methods, and
  applications.
\newblock {\em arXiv preprint arXiv:1901.04592\/}.

\bibitem[\protect\citeauthoryear{Plumb, Molitor, and Talwalkar}{Plumb
  et~al.}{2018}]{plumb2018model}
Plumb, G., D.~Molitor, and A.~Talwalkar (2018).
\newblock Model agnostic supervised local explanations.
\newblock {\em arXiv preprint arXiv:1807.02910\/}.

\bibitem[\protect\citeauthoryear{Prosperi, Guo, Sperrin, Koopman, Min, He,
  Rich, Wang, Buchan, and Bian}{Prosperi et~al.}{2020}]{prosperi2020causal}
Prosperi, M., Y.~Guo, M.~Sperrin, J.~S. Koopman, J.~S. Min, X.~He, S.~Rich,
  M.~Wang, I.~E. Buchan, and J.~Bian (2020).
\newblock Causal inference and counterfactual prediction in machine learning
  for actionable healthcare.
\newblock {\em Nature Machine Intelligence\/}~{\em 2\/}(7), 369--375.

\bibitem[\protect\citeauthoryear{{Protection Regulation}}{{Protection
  Regulation}}{2018}]{regulation2018general}
{Protection Regulation} (2018).
\newblock General data protection regulation.
\newblock {\em Intouch\/}.

\bibitem[\protect\citeauthoryear{Selesnick}{Selesnick}{2017}]{selesnick2017sparse}
Selesnick, I. (2017).
\newblock Sparse regularization via convex analysis.
\newblock {\em IEEE Transactions on Signal Processing\/}~{\em 65\/}(17),
  4481--4494.

\bibitem[\protect\citeauthoryear{Shukla and Fricklas}{Shukla and
  Fricklas}{2018}]{shukla2018machine}
Shukla, N. and K.~Fricklas (2018).
\newblock {\em Machine learning with TensorFlow}.
\newblock Manning Greenwich.

\bibitem[\protect\citeauthoryear{{S\&P Global}}{{S\&P
  Global}}{2018}]{ratings2018guide}
{S\&P Global} (2018).
\newblock Guide to credit rating essentials: What are credit ratings and how do
  they work.

\bibitem[\protect\citeauthoryear{{Standard and Poor's Corporation}}{{Standard
  and Poor's Corporation}}{1981}]{spguidecr}
{Standard and Poor's Corporation} (1981).
\newblock {\em Standard \& Poor's Guide to Credit Rating Essentials}.
\newblock Standard \& Poor's.

\bibitem[\protect\citeauthoryear{Wachter, Mittelstadt, and Russell}{Wachter
  et~al.}{2017}]{wachter2017counterfactual}
Wachter, S., B.~Mittelstadt, and C.~Russell (2017).
\newblock Counterfactual explanations without opening the black box: Automated
  decisions and the gdpr.
\newblock {\em Harv. JL \& Tech.\/}~{\em 31}, 841.

\bibitem[\protect\citeauthoryear{Wallis, Kumar, and Gepp}{Wallis
  et~al.}{2019}]{wallis2019credit}
Wallis, M., K.~Kumar, and A.~Gepp (2019).
\newblock Credit rating forecasting using machine learning techniques.
\newblock In {\em Managerial Perspectives on Intelligent Big Data Analytics},
  pp.\  180--198. IGI Global.

\bibitem[\protect\citeauthoryear{Wang, Wang, and Florescu}{Wang
  et~al.}{2020}]{wang2020image}
Wang, D., T.~Wang, and I.~Florescu (2020).
\newblock Is image encoding beneficial for deep learning in finance?
\newblock {\em IEEE Internet of Things Journal\/}.

\bibitem[\protect\citeauthoryear{West}{West}{2000}]{west2000neural}
West, D. (2000).
\newblock Neural network credit scoring models.
\newblock {\em Computers \& Operations Research\/}~{\em 27\/}(11-12),
  1131--1152.

\bibitem[\protect\citeauthoryear{Yang, Yang, Dyer, He, Smola, and Hovy}{Yang
  et~al.}{2016}]{yang2016hierarchical}
Yang, Z., D.~Yang, C.~Dyer, X.~He, A.~Smola, and E.~Hovy (2016).
\newblock Hierarchical attention networks for document classification.
\newblock In {\em Proceedings of the 2016 conference of the North American
  chapter of the association for computational linguistics: human language
  technologies}, pp.\  1480--1489.

\bibitem[\protect\citeauthoryear{Ye, Liu, and Li}{Ye
  et~al.}{2008}]{ye2008multiclass}
Ye, Y., S.~Liu, and J.~Li (2008).
\newblock A multiclass machine learning approach to credit rating prediction.
\newblock In {\em 2008 International Symposiums on Information Processing},
  pp.\  57--61. IEEE.

\bibitem[\protect\citeauthoryear{Yuan and Ghanem}{Yuan and
  Ghanem}{2016}]{yuan2016sparsity}
Yuan, G. and B.~Ghanem (2016).
\newblock Sparsity constrained minimization via mathematical programming with
  equilibrium constraints.
\newblock {\em arXiv preprint arXiv:1608.04430\/}.

\bibitem[\protect\citeauthoryear{Zhang, Fan, Wang, Zhou, and Tao}{Zhang
  et~al.}{2020}]{zhang2020top}
Zhang, X., M.~Fan, D.~Wang, P.~Zhou, and D.~Tao (2020).
\newblock Top-k feature selection framework using robust 0-1 integer
  programming.
\newblock {\em IEEE Transactions on Neural Networks and Learning Systems\/}.

\bibitem[\protect\citeauthoryear{Zhao, Xu, Kang, Kabir, Liu, and Wasinger}{Zhao
  et~al.}{2015}]{zhao2015investigation}
Zhao, Z., S.~Xu, B.~H. Kang, M.~M.~J. Kabir, Y.~Liu, and R.~Wasinger (2015).
\newblock Investigation and improvement of multi-layer perceptron neural
  networks for credit scoring.
\newblock {\em Expert Systems with Applications\/}~{\em 42\/}(7), 3508--3516.

\end{thebibliography}
\newpage
\section{Appendix}\label{sec:appendix}

\begin{table}[!htbp]
  \centering
  \caption{A complete list of accounting variables that may not be changed}\label{tab:all_fea_not_manage}
  \begin{adjustbox}{width=\columnwidth,center}
    \begin{tabular}{p{12.5em}|p{58em}}
    \toprule
    \textbf{Reasons} & \textbf{Features} \\
    \midrule
    Adjustment, scheduled items & Accounting Changes - Cumulative Effect, Accumulated Other Comprehensive Income (Loss), Assets Netting \& Other Adjustments, Accum Other Comp Inc - Other Adjustments, Accum Other Comp Inc - Min Pension Liab Adj, Comp Inc - Beginning Net Income, Comp Inc - Currency Trans Adj, Comp Inc - Other Adj, Comp Inc - Minimum Pension Adj, Dilution Adjustment, Accum Other Comp Inc - Marketable Security Adjustments, Provision for Loan/Asset Losses, Pension Core Adjustment - 12mm, Core Pension Adjustment Diluted EPS Effect 12MM, Core Pension Adjustment Diluted EPS Effect, Core Pension Adjustment Basic EPS Effect 12MM, Core Pension Adjustment Basic EPS Effect, Core Pension Interest Adjustment After-tax Preliminary, Core Pension Interest Adjustment After-tax, Core Pension Interest Adjustment Diluted EPS Effect Preliminary, Core Pension Interest Adjustment Diluted EPS Effect, Core Pension Interest Adjustment Basic EPS Effect Preliminary, Core Pension Interest Adjustment Basic EPS Effect, Core Pension Interest Adjustment Pretax Preliminary, Core Pension Interest Adjustment Pretax, Core Pension Adjustment 12MM Diluted EPS Effect Preliminary, Core Pension Adjustment Diluted EPS Effect Preliminary, Core Pension Adjustment 12MM Basic EPS Effect Preliminary, Core Pension Adjustment Basic EPS Effect Preliminary, Core Pension Adjustment Preliminary, Core Pension Adjustment, Core Pension w/o Interest Adjustment After-tax Preliminary, Core Pension w/o Interest Adjustment After-tax, Core Pension w/o Interest Adjustment Diluted EPS Effect Preliminary, Core Pension w/o Interest Adjustment Diluted EPS Effect, Core Pension w/o Interest Adjustment Basic EPS Effect Preliminary, Core Pension w/o Interest Adjustment Basic EPS Effect, Core Pension w/o Interest Adjustment Pretax Preliminary, Core Pension w/o Interest Adjustment Pretax, Core Post Retirement Adjustment, Core Post Retirement Adjustment Diluted EPS Effect 12MM, Core Post Retirement Adjustment Diluted EPS Effect, Core Post Retirement Adjustment 12MM, Core Post Retirement Adjustment Basic EPS Effect 12MM, Core Post Retirement Adjustment Basic EPS Effect, Core Post Retirement Adjustment 12MM Diluted EPS Effect Preliminary, Core Post Retirement Adjustment Diluted EPS Effect Preliminary, Core Post Retirement Adjustment 12MM Basic EPS Effect Preliminary, Core Post Retirement Adjustment Basic EPS Effect Preliminary, Core Post Retirement Adjustment Preliminary, Receivables - Estimated Doubtful, Accum Other Comp Inc - Cumulative Translation Adjustments, Reserve for Loan/Asset Losses \\
    % \multicolumn{1}{l|}{} & \multicolumn{1}{l}{} \\
    % \multicolumn{1}{l|}{} & \multicolumn{1}{l}{} \\
    % \multicolumn{1}{l|}{} & \multicolumn{1}{l}{} \\
    % \multicolumn{1}{l|}{} & \multicolumn{1}{l}{} \\
    % \multicolumn{1}{l|}{} & \multicolumn{1}{l}{} \\
    % \multicolumn{1}{l|}{} & \multicolumn{1}{l}{} \\
    \midrule
    Special items, unusual or non-recurring items & Acquisition/Merger After-Tax, Acquisition/Merger Diluted EPS Effect, Acquisition/Merger Basic EPS Effect, Acquisition/Merger Pretax, Extinguishment of Debt After-tax, Extinguishment of Debt Diluted EPS Effect, Extinguishment of Debt Basic EPS Effect, Extinguishment of Debt Pretax, Impairments of Goodwill AfterTax - 12mm, Impairment of Goodwill After-tax, Impairments Diluted EPS - 12mm, Impairment of Goodwill Diluted EPS Effect, Impairment of Goodwill Basic EPS Effect 12MM, Impairment of Goodwill Basic EPS Effect, Impairment of Goodwill Pretax, Gain/Loss After-Tax, Gain/Loss on Sale (Core Earnings Adjusted) After-tax 12MM, Gain/Loss on Sale (Core Earnings Adjusted) After-tax, Gain/Loss on Sale (Core Earnings Adjusted) Diluted EPS Effect 12MM, Gain/Loss on Sale (Core Earnings Adjusted) Diluted EPS, Gain/Loss on Sale (Core Earnings Adjusted) Basic EPS Effect 12MM, Gain/Loss on Sale (Core Earnings Adjusted) Basic EPS Effect, Gain/Loss on Sale (Core Earnings Adjusted) Pretax, Gain/Loss Diluted EPS Effect, Gain/Loss Basic EPS Effect, Gain/Loss Pretax, Gain/Loss on Ineffective Hedges, Inventory - Other, Nonperforming Assets - Total, Nonrecurring Income Taxes Diluted EPS Effect, Nonrecurring Income Taxes Basic EPS Effect, Nonrecurring Income Taxes - After-tax, Order backlog, Restructuring Cost After-tax, Restructuring Cost Diluted EPS Effect, Restructuring Cost Basic EPS Effect, Restructuring Cost Pretax, Other Special Items Diluted EPS Effect, Other Special Items Basic EPS Effect, Other Special Items After-tax, Other Special Items Pretax, Special Items, Writedowns After-tax, Writedowns Diluted EPS Effect, Writedowns Basic EPS Effect, Writedowns Pretax \\
    \midrule
    Assets are discontinued operations & Other Long-term Assets, Discontinued Operations, Extraordinary Items and Discontinued Operations \\
    \midrule
    Assets are in the Market & Accum Other Comp Inc - Derivatives Unrealized Gain/Loss, Accum Other Comp Inc - Unreal G/L Ret Int in Sec Assets, Assets Level2 (Observable), Comp Inc - Derivative Gains/Losses, Comp Inc - Securities Gains/Losses, Common Shares Used to Calculate Earnings Per Share - 12 Months Moving, Com Shares for Diluted EPS, Common Shares Issued, Common/Ordinary Stock (Capital), Dividends - Preferred/Preference, Earnings Per Share (Diluted) - Including Extraordinary Items, Earnings Per Share (Diluted) - Excluding Extraordinary items, Earnings Per Share (Basic) - Including Extraordinary Items, Earnings Per Share (Basic) - Excluding Extraordinary Items, Earnings Per Share (Basic) - Excluding Extraordinary Items - 12 Months Moving, Foreign Exchange Income (Loss), Goodwill (net), Options - Fair Value of Options Granted, Life of Options - Assumption (\# yrs), Risk Free Rate - Assumption (\%), Volatility - Assumption (\%), Repurchase Price - Average per share Quarter, Preferred/Preference Stock - Nonredeemable, Preferred/Preference Stock (Capital) - Total, Preferred/Preference Stock - Redeemable, Implied Option Expense - 12mm, Implied Option EPS Diluted 12MM, Implied Option 12MM EPS Diluted Preliminary, Implied Option EPS Diluted, Implied Option EPS Diluted Preliminary, Implied Option EPS Basic 12MM, Implied Option 12MM EPS Basic Preliminary, Implied Option EPS Basic, Implied Option EPS Basic Preliminary, Implied Option Expense, Implied Option Expense Preliminary \\
    \midrule
    Accord, regulated items & Risk-Adjusted Capital Ratio - Tier 1, Risk-Adjusted Capital Ratio - Tier 2, Risk-Adjusted Capital Ratio - Combined \\
    \midrule
    Agreements with shareholders, employees & Total Shares Repurchased - Quarter, Common Shares Outstanding, Common Shares Used to Calculate Earnings Per Share - Basic, Carrying Value, Common Stock Equivalents - Dollar Savings, Deferred Compensation, Dividends - Preferred/Preference, Common ESOP Obligation - Total, Preferred ESOP Obligation - Non-Redeemable, Preferred ESOP Obligation - Redeemable, Preferred ESOP Obligation - Total, Dividend Rate - Assumption (\%), Nonred Pfd Shares Outs (000) - Quarterly, Redeem Pfd Shares Outs (000), Other Stockholders- Equity Adjustments, Stock Compensation Expense, Treasury Stock - Number of Common Shares, Treasury Stock - Total (All Capital) \\
    \midrule
    Computational Items & Accumulated Depreciation of RE Property, Depreciation, Depletion and Amortization (Accumulated), Depreciation and Amortization - Total, Depr/Amort of Property, Amortization of Goodwill, Receivables - Current Other incl Tax Refunds, Total Fair Value Changes including Earnings, Total Fair Value Liabilities, Deferred Tax Asset - Long Term, Current Deferred Tax Asset, Current Deferred Tax Liability, Deferred Taxes - Balance Sheet, Income Taxes - Deferred, Deferred Taxes and Investment Tax Credit, Income Taxes Payable, Income Taxes - Total, Excise Taxes \\
    \midrule
    Special events & Reversal - Restructruring/Acquisition Aftertax 12MM, Reversal - Restructruring/Acquisition Aftertax, Reversal - Restructuring/Acq Diluted EPS Effect 12MM, Reversal - Restructuring/Acq Diluted EPS Effect, Reversal - Restructuring/Acq Basic EPS Effect 12MM, Reversal - Restructuring/Acq Basic EPS Effect, Reversal - Restructruring/Acquisition Pretax, Settlement (Litigation/Insurance) AfterTax - 12mm, Settlement (Litigation/Insurance) After-tax, Settlement (Litigation/Insurance) Diluted EPS Effect 12MM, Settlement (Litigation/Insurance) Diluted EPS Effect, Settlement (Litigation/Insurance) Basic EPS Effect 12MM, Settlement (Litigation/Insurance) Basic EPS Effect, Settlement (Litigation/Insurance) Pretax, Extraordinary Items \\
    \midrule
    Noncontrolling loss/gain from subsidiary & Comprehensive Income - Noncontrolling Interest, Equity in Earnings (I/S) - Unconsolidated Subsidiaries, Investment and Advances - Equity, Investment and Advances - Other, Noncontrolling Interests - Nonredeemable - Balance Sheet, Noncontrolling Interest - Redeemable - Balance Sheet, Noncontrolling Interests - Total - Balance Sheet, Noncontrolling Interest - Income Account \\
    \midrule
    Non-operating items & Non-Operating Income (Expense) - Total, Gain/Loss on Sale of Property \\
    \bottomrule
    \end{tabular}%
    \end{adjustbox}
\end{table}%

\end{document}